\documentclass[a4paper,twocolumn,10pt, accepted=2026-03-16]{quantumarticle}
\pdfoutput=1

\usepackage[utf8]{inputenc}
\usepackage[english]{babel}
\usepackage[T1]{fontenc}
\usepackage[numbers,sort&compress]{natbib}
\usepackage{listings}

\usepackage{graphicx}
\usepackage{amsmath, amssymb}
\usepackage{braket}
\usepackage{bbm}
\usepackage{subcaption}
\usepackage{ragged2e}
\usepackage{xcolor}
\usepackage{hyperref}

\newcommand{\ustuttgartIII}{Institute for Theoretical Physics III and Center for Integrated Quantum Science and Technology,
University of Stuttgart, 70550 Stuttgart, Germany}
\newcommand{\ustuttgartV}{5th Institute for Physics and Center for Integrated Quantum Science and Technology,
University of Stuttgart, 70550 Stuttgart, Germany}

\begin{document}

\title{Suppressing crosstalk for Rydberg quantum gates}

\author{Gina Warttmann}
\affiliation{\ustuttgartIII}
\orcid{0009-0002-3069-6945}

\author{Florian Meinert}
\affiliation{\ustuttgartV}
\orcid{0000-0002-9106-3001}

\author{Hans Peter Büchler}
\affiliation{\ustuttgartIII}
\orcid{0000-0002-7233-6828}

\author{Sebastian Weber}
\affiliation{\ustuttgartIII}
\orcid{0000-0001-9763-9131}

\begin{abstract}
We present a method to suppress crosstalk from implementing controlled-Z gates via local addressing in neutral atom quantum computers. 
In these systems, a fraction of the laser light that is applied locally to implement gates typically leaks to other atoms.
We analyze the resulting crosstalk in a setup of two gate atoms and one neighboring third atom. We then perturbatively derive a spin-echo-inspired gate protocol that suppresses the leading order of the amplitude error, which dominates the crosstalk. Numerical simulations demonstrate that our gate protocol improves the fidelity by two orders of magnitude across a broad range of experimentally relevant parameters. To further reduce the infidelity, we develop a circuit to cancel remaining phase errors. Our results pave the way for using local addressing for high-fidelity quantum gates on Rydberg-based quantum computers.
\end{abstract}

\maketitle

\section{\label{sec: introduction} Introduction}
Neutral atoms trapped in optical tweezers or lattices are promising for the realization of universal quantum computers \cite{henriet_2020, saffman_2010, wu_2021}. On the neutral-atom platform, qubit states are encoded in long-lived electronic states of atoms that can be controlled by laser pulses. As of now, arrays of more than a thousand trapped atoms have been realized \cite{huft_2022, pause_2024, manetsch_2024, pichard_2024,  norcia_2024}. 
To implement two- and multi-qubit gates between the atoms, the atoms are temporarily excited to strongly interacting Rydberg states \cite{jaksch_2000, graham_2022, saffman_2016, levine_2019,  pagano_2022, jandura_2022}. By tuning the excitation laser pulse, controlled phase gates have been implemented with fidelities above $99.7\%$ \cite{muniz_2025, tsai_2025}. 
However, addressing of individual atoms remains a major challenge, and the high fidelities have been reached only in setups where the gate atoms were moved away from other atoms prior to performing a gate. 
In setups where the gate atoms are not spatially separated from other atoms, some fraction of the excitation laser light typically leaks to other atoms  \cite{saffman_2016, radnaev_2025}. 
Here, we analyze the resulting crosstalk for a controlled-Z gate and develop a gate protocol to suppress it. 

The ability to perform entangling gates on individual qubits is needed for gate-based universal quantum computing \cite{barenco_1995, Nielsen_Chuang_2010}. For the neutral-atom platform, two complementary methods exist for the execution of entangling gates on individual atomic qubits: The first method makes use of the ability to shift the atoms in space \cite{bluvstein_2022, barnes_2022}. This allows for bringing two gate atoms close to each other, while having a large distance to other atoms. After this rearrangement step, an excitation laser pulse is applied globally to all atoms. Thanks to the finite range of the Rydberg interaction, atoms that are not close to other atoms undergo a trivial rotation. Note that by bringing multiple atoms pairwise close to each other, multiple entangling gates can be performed in parallel (field programmable qubit array, FPQA, architecture \cite{bluvstein_2022, tan_2024, wang_2024, tan_25, muniz_2025, wang_2024_2}).
As an alternative to a truly global excitation laser, one can use a laser that is focused on an interaction zone (interaction-zone architecture \cite{barnes_2022, reichard_2025, muniz_2025_2}). 
Outside this zone, the atoms can be stored in a dense register. 
To perform entangling gates, atoms are brought pairwise to the interaction zone. 
The alternative method is to use local addressing where a laser is directly focused down on the gate atoms \cite{burgers_2022, isenhower_2010, radnaev_2025, graham_2019}. 
This has the benefit that shift operations are not needed. 
Because shift operations with a typical speed of about $1\; \mu \text{m} / \mu \text{s}$ are much slower than gate times of a few hundred nanoseconds \cite{bluvstein_2022}, this can reduce idling errors. Moreover, local addressing allows for packing atoms in denser arrays than in the FPQA architecture and for a better parallelization of gate operations than in the interaction-zone architecture. However, local addressing is technically challenging as it requires micrometer-sized laser spots. 
This is because the separation of the atoms is typically a few micrometers due to the finite range of the Rydberg interaction. For these small spacings, it is unavoidable that some fraction of the laser light leaks to other atoms. 
In addition to technical limitations such as aberrations and scattering from the optical components \cite{sotirova_2024, radnaev_2025}, a finite numerical aperture leads to deviations from an ideal Gaussian beam profile \cite{tanaka_1985}. 

In this paper, we analyze the crosstalk that results from the laser light leaking to other atoms and develop a gate protocol to suppress it. In Sec. \ref{sec: cz gate}, we introduce a setup for studying the crosstalk. Our setup comprises two gate atoms on which a controlled-Z gate is performed and a neighboring third atom that is subject to a fraction of the laser light. In Sec. \ref{sec: pert. description }, we use perturbation theory to analyze the effect of the crosstalk and develop a gate protocol to suppress it in leading order. In Secs. \ref{sec: volle numerik} and \ref{sec: vdw variieren}, we numerically simulate our gate protocol, providing a reduction of the infidelity of two orders of magnitude over a broad range of experimentally relevant parameters. In Sec. \ref{sec: phasen variieren}, we develop a circuit for canceling remaining phase errors, which reduces the infidelity by another order of magnitude.

\section{\label{sec: cz gate}Setup}
We consider a realization of a controlled-Z gate with local addressing where the excitation laser is focused down on the two gate atoms, but a non-zero intensity also affects surrounding atoms.
We examine the resulting crosstalk using a setup of two gate atoms and a third neighboring atom with Hamiltonian
\begin{equation}
    \label{eq: general H}
    H = H_{\scriptscriptstyle \mathrm{cz}} + H_{\scriptscriptstyle 3} + H_{\scriptscriptstyle \mathrm{int}}.
\end{equation}
The dynamics of the gate atoms is captured by $H_{\scriptscriptstyle \mathrm{cz}}$, whereas $H_{\scriptscriptstyle 3}$ describes the neighboring third atom. 
The interaction between the third atom and the gate atoms is characterized by $H_{\scriptscriptstyle \mathrm{int}}$. 
\begin{figure}[t]
    \centering
    \includegraphics[width=0.407\textwidth]{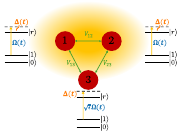}
\caption{\justifying Setup for studying the
crosstalk of a two-qubit gate (on atoms 1 and 2) affecting a third neighboring atom (atom 3). 
Each atom is modeled as a
three-level system with states $\ket{0}, \ket{1}$, and a Rydberg state $\ket{r}$. 
A local laser (depicted in yellow) couples $\ket{1}$ to $\ket{r}$ with Rabi frequency $\Omega(t)$ and detuning $\Delta(t)$. 
We assume that the third atom is affected by a small fraction $\sqrt{\epsilon}\Omega(t)$ of the Rabi frequency.
Atoms in the Rydberg state interact via van der Waals interaction $V_{\scriptscriptstyle ij}$ (green arrows).}
    \label{fig: setup sketch}
\end{figure}
Each of the three atoms comprises the qubit states $\ket{0}$ and $\ket{1}$, encoded in long-lived electronic states, and a Rydberg state $\ket{r}$.
For simplicity, we first consider an equilateral arrangement as depicted in Fig. \ref{fig: setup sketch}, where all three atoms are separated by the same distance. 
In Sec. \ref{sec: vdw variieren}, we will generalize this configuration by varying the interaction strength between the third atom and the gate atoms.

The controlled-Z gate is implemented on atoms 1 and 2 as in Ref. \cite{pagano_2022}.
We couple state $\ket{1}_{\scriptscriptstyle i}$ of the $i^\text{th}$ atom to the strongly interacting Rydberg state $\ket{r}_{\scriptscriptstyle i}$ with a time-dependent Rabi frequency $\Omega(t)$ and a time-dependent detuning $\Delta(t)$ as described by the Hamiltonian 
\begin{equation}
    \label{eq: H gate}
    H_{\scriptscriptstyle \mathrm{cz}} = \hbar \sum_{i=1}^{2}\left[\frac{\Omega(t)}{2}\left(\sigma_{\scriptscriptstyle i}^+ + \sigma_{\scriptscriptstyle i}^-\right) - \Delta(t)n_{\scriptscriptstyle i}\right] + V_{\scriptscriptstyle 12}n_{\scriptscriptstyle 1}n_{\scriptscriptstyle 2}
\end{equation}
with $\sigma_{\scriptscriptstyle i}^+=\ket{r}\bra{1}_{\scriptscriptstyle i}$, $\sigma_{\scriptscriptstyle i}^-=\ket{1}\bra{r}_{\scriptscriptstyle i}$  and $n_{\scriptscriptstyle i} = \ket{r}\bra{r}{\scriptscriptstyle i}$. 
The two atoms interact by van der Waals interaction of strength $V_{\scriptscriptstyle 12}$ between the Rydberg levels.
To realize the two-qubit controlled-Z gate, we aim to operate in the regime of strong Rydberg blockade $V_{\scriptscriptstyle ij}/\hbar\Omega_{\scriptscriptstyle 0}\gg 1$ with a maximal Rabi frequency $\Omega_{\scriptscriptstyle 0}$. 
By tuning the pulse shape $\Omega(t)$ and $\Delta(t)$, \cite{pagano_2022, levine_2019, jandura_2022}, one can achieve that the dynamics of the gate atoms under $H_{\scriptscriptstyle \mathrm{cz}}$ can be described by the mapping
\begin{equation}
\label{eq: mapping CZ gate}
    \begin{split}
        &\ket{00} \rightarrow \ket{00}\\
        &\ket{01} \rightarrow \ket{01} e^{i\varphi}\\
        &\ket{10} \rightarrow \ket{10} e^{i\varphi}\\
        &\ket{11} \rightarrow \ket{11} e^{i(2\varphi-\pi)}.\\
    \end{split}
\end{equation}
Due to the symmetric setup, initial gate states $\ket{01}$ and $\ket{10}$ evolve identically.
If the two gate atoms are initialized in state $\ket{00}$, they are not affected by $H_{\scriptscriptstyle \mathrm{cz}}$. 

In the following, we examine two different pulse protocols.
The single-pulse protocol corresponds to the pulse shape found in Ref. \cite{pagano_2022} and is depicted in Fig. \ref{fig: setup}(a). 
As ansatz functions, a Gaussian detuning sweep and a smoothly turned on and off Rabi frequency, $\Omega(t)/\Omega_0 \propto \tanh(t/\kappa)$, are considered. 
We use the rise time $\kappa\Omega_0 = 0.31$ and $4.5 w$ as the duration of the Rabi frequency sweep, where $w$ is the width of the Gaussian. The width, the amplitude, and the offset of the Gaussian detuning sweep are optimized to obtain a gate fulfilling Eq. \eqref{eq: mapping CZ gate}. 
We consider an experimentally feasible van der Waals interaction strength of $V_{\scriptscriptstyle 12}/\hbar\Omega_{\scriptscriptstyle 0} = 21.1$ \cite{pagano_2022}.

To reduce crosstalk from the laser light leaking to the third atom, we split the single pulse into two pulses, each implementing a controlled-$\pi/2$ gate. 
This double-pulse protocol is shown in Fig. \ref{fig: setup}(c) and discussed in the next section.

The intensity of the laser light leaking to the third atom is considered to be a fraction $\epsilon$ of the intensity $I \propto \Omega^2$ of the local addressing beam.
Hence, the dynamics of the third atom are governed by 
\begin{equation}
    \label{eq: H 3rd}
    H_{\scriptscriptstyle 3} =\frac{\sqrt{\epsilon}\Omega(t)\hbar}{2}\left(\sigma_{\scriptscriptstyle 3}^+ + \sigma_{\scriptscriptstyle 3}^-\right) - \underbrace{\Delta(t) n_{\scriptscriptstyle 3}\hbar}_{\substack{H_{\scriptscriptstyle 0}}}.
\end{equation}
The interaction between the gate atoms and the third one is described by 
\begin{equation}
    \label{eq: H int 13 23}
    H_{\scriptscriptstyle \mathrm{int}} = \sum_{i=1}^{2}V_{\scriptscriptstyle i3}n_{\scriptscriptstyle i}n_{\scriptscriptstyle 3},
\end{equation}
where $V_{\scriptscriptstyle i3}$ characterizes the van der Waals interaction between gate atom $i$ and the third one.

\section{\label{sec: pert. description } Perturbative Description }
\begin{figure}[t]
    \centering
    \includegraphics[width=0.48\textwidth]{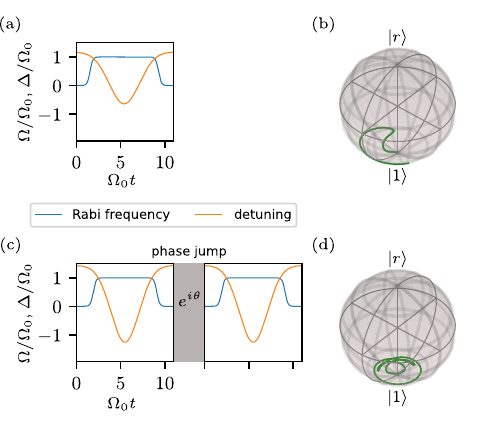}
\caption{\justifying Comparison of the single- and double-pulse protocol.
(a) The controlled-Z gate can be realized by a single pulse with a Gaussian detuning sweep $\Delta(t)$ and a smoothly turned on and off Rabi frequency $\Omega(t)$  \cite{pagano_2022}.
(b) The green line shows the evolution of the state of the third atom on the Bloch sphere during the single-pulse protocol if initialized in $\ket{1}$.
After the pulse, the probability of being in the Rydberg state $\ket{r}$ remains finite.
(c) To suppress the crosstalk, we introduce the double-pulse protocol. 
The gate is split into two controlled-$\pi/2$ gates, realizing a controlled-Z gate on the gate atoms. Between the gate pulses, the phase of the excitation laser is changed so that 
on the third atom, the second pulse brings the Rydberg population back to $\ket{1}$ (d).}
\label{fig: setup}
\end{figure}

In this section, we use a perturbative approach to describe the time evolution of the third atom when driving the controlled-Z gate. 
For our approach to be valid, $H_{\scriptscriptstyle \mathrm{int}}$ must be negligible.
This requirement is met if the gate is initially prepared in the state $\ket{00}$. 
In this case, gate and third atom evolve independently. 
For initial gate states $\ket{10}$ and $\ket{11}$, excitations of the third atom are suppressed by the strong Rydberg blockade. 
This means the initial gate state $\ket{00}$ is the most relevant state for our discussion. 
Assuming that the influence of the Rabi frequency $\sqrt{\epsilon}\Omega(t)$ on the third atom is only a small perturbation compared to $H_{\scriptscriptstyle 0}$, we can apply perturbation theory in the parameter $\sqrt{\epsilon}$.
The time evolution operator of the unperturbed Hamiltonian $H_{\scriptscriptstyle 0}$ reads
\begin{equation}
    \label{eq: U0}
    U_{\scriptscriptstyle 0}(t, t_{\scriptscriptstyle 0}) = e^{i\phi(t, t_{\scriptscriptstyle 0})\ket{r}\bra{r}}, 
\end{equation}
with
\begin{equation}
    \label{eq: phi func}
    \phi(t, t_{\scriptscriptstyle 0}) = \int_{t_{\scriptscriptstyle 0}}^t\text{d}t' \Delta(t').
\end{equation}
Expanding the time evolution operator  of the perturbed Hamiltonian $H_{\scriptscriptstyle 3}$ yields in the interaction picture
\begin{widetext}
\begin{equation}
\begin{split}
    \label{eq: U tilde}
    U(t, t_{\scriptscriptstyle 0}) &= \mathbbm{1} - \frac{i\sqrt{\epsilon}}{2}\int_{t_{\scriptscriptstyle 0}}^t \text{d}t_{\scriptscriptstyle 1} \Omega(t_{\scriptscriptstyle 1})\left( e^{-i\phi(t_{\scriptscriptstyle 1}, t_{\scriptscriptstyle 0})}\sigma^+ + e^{i\phi(t_{\scriptscriptstyle 1}, t_{\scriptscriptstyle 0})}\sigma^- \right)\\
    &- \frac{\epsilon}{4}\int_{t_{\scriptscriptstyle 0}}^t \text{d}t_{\scriptscriptstyle 1} \int_{t_{\scriptscriptstyle 0}}^{t_{\scriptscriptstyle 1}} \text{d}t_{\scriptscriptstyle 2} \Omega(t_{\scriptscriptstyle 1})\Omega(t_{\scriptscriptstyle 2})\left( e^{-i\left(\phi(t_{\scriptscriptstyle 1}, t_{\scriptscriptstyle 0}) - \phi(t_{\scriptscriptstyle 2}, t_{\scriptscriptstyle 0}) \right)}\ket{r}\bra{r} 
    + e^{i\left(\phi(t_{\scriptscriptstyle 1}, t_{\scriptscriptstyle 0}) - \phi(t_{\scriptscriptstyle 2}, t_{\scriptscriptstyle 0}) \right)}\ket{1}\bra{1} \right)\\
    &+ \mathcal{O}\left(\sqrt{\epsilon}^3\right).
\end{split}
\end{equation}
\end{widetext}
If initialized in $\ket{1}$, the third atom can be excited to the Rydberg state $\ket{r}$ and/or collect a phase:
\begin{equation}
    \label{eq: third in 1}
    U(t_{\scriptscriptstyle f}, t_{\scriptscriptstyle 0}) \ket{1} = \sqrt{\left(1-p(\epsilon)\right)}e^{i\varphi(\epsilon)}  \ket{1} + \sqrt{p(\epsilon) }e^{i\varphi'(\epsilon)}\ket{r}.
\end{equation}
Here, the amplitude error, i.e., the probability of the third atom being excited to the Rydberg state is
\begin{equation}
    \label{eq: amplitude error} 
    p(\epsilon) = \left|\bra{r} U(t_{\scriptscriptstyle f}, t_{\scriptscriptstyle 0})\ket{1}\right|^2 = \alpha \epsilon + \mathcal{O}(\epsilon^2), ~~~ \alpha \in \mathbbm{R}.
\end{equation}
The phase of the state $\ket{1}$ picked up during the laser pulse is the phase error
\begin{equation}
    \label{eq: phase error}
    \varphi(\epsilon) = \text{arg}\left(\bra{1}U(t_{\scriptscriptstyle f}, t_{\scriptscriptstyle 0})\ket{1}\right) = \beta\epsilon + \mathcal{O}(\epsilon^2), ~~~ \beta \in \mathbbm{R}.
\end{equation}
The phase $\varphi'(\epsilon)$ of the Rydberg state will not be of relevance for our analysis.
To quantify the crosstalk, we introduce the fidelity 
\begin{equation}
    \label{eq: fid of 3rd qubit}
    \mathcal{F}_{\scriptscriptstyle 3} = \left| \braket{\Psi_{\scriptscriptstyle 0}|\Psi{\scriptscriptstyle f}}\right|^2,
\end{equation}
for the third atom to stay in $\ket{\Psi_{\scriptscriptstyle 0}} = 1/\sqrt{2}\left(\ket{0} + \ket{1}\right)$ as a measure that is sensitive to both errors.
Notably, up to linear order in $\epsilon$, the fidelity $\mathcal{F}_{\scriptscriptstyle 3} \approx 1 - \frac{\alpha}{2}\epsilon$ depends on the amplitude error only. 
Hence, to achieve fidelities close to one, we aim at suppressing the amplitude error.  
To gain intuition about how this can be achieved, we take a look at the path on the Bloch sphere. 
Applying the single-pulse protocol to realize the controlled-Z gate in the case where the third atom is initialized in $\ket{1}$, there is a non-negligible probability that it will be excited to the Rydberg state.
This can be seen as the path on the Bloch sphere, shown in Fig. \ref{fig: setup}(b), is not closed.
Hence, we replace the single-pulse protocol with a pulse protocol consisting of two pulses as depicted in Fig. \ref{fig: setup}(d) such that each performs a controlled-$\pi/2$ gate on the gate atoms; thus, in total, a controlled-Z gate.
On the third atom, the second pulse brings the population in $\ket{r}$ back to the qubit state $\ket{1}$, see Fig. \ref{fig: setup}(d). 
This requires a phase jump between both pulses
\begin{equation}
    \label{eq: Rabi freq 2nd pulse}
    \Omega(t) \rightarrow \Omega(t)e^{i\theta}
\end{equation}
with $\theta = \phi \left(t_{\scriptscriptstyle f}/2, t_{\scriptscriptstyle 0} \right) + \pi$ such that the amplitude error vanishes in leading order at the end of the protocol.
The perturbative approach predicts the infidelity
$1-\mathcal{F}_{\scriptscriptstyle 3} = 6.91\epsilon^2 + \mathcal{O}(\epsilon^4)$, suppressing the linear order in $\epsilon$. 

\section{\label{sec: volle numerik} Full Simulation}
\begin{figure}[t]
\centering
    \begin{subfigure}{0.48\textwidth}
        \includegraphics[width=1\linewidth]{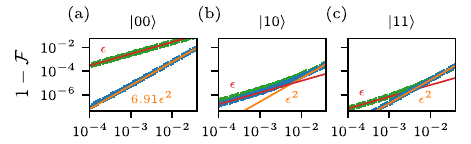}
    \end{subfigure}
    \begin{subfigure}{0.465\textwidth}
        \includegraphics[width=1\linewidth]{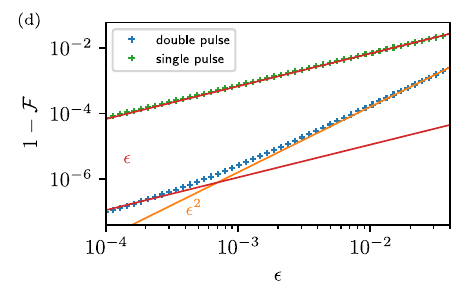}
    \end{subfigure}
\caption{\justifying{ Comparison of the infidelity $1-\mathcal{F}$ of the
effective three-qubit gate after the single- and double-pulse protocols for the equilateral setup with $V_{\scriptscriptstyle 13}/\hbar\Omega_{\scriptscriptstyle 0} = V_{\scriptscriptstyle 23}/\hbar\Omega_{\scriptscriptstyle 0} = V_{\scriptscriptstyle 12}/\hbar\Omega_{\scriptscriptstyle 0} = 21.1$ \cite{pagano_2022}. 
Infidelity when the gate atoms were initialized in states (a) $\ket{00}$, (b) $\ket{10}$}, and (c) $\ket{11}$, and the third atom in $1/\sqrt{2}\left(\ket{0}+\ket{1}\right)$. 
For $\ket{00}$, the numerical data for the double-pulse protocol (blue crosses) agrees very well with the predicted $\epsilon^2$-increase (orange line). The linear dependence on $\epsilon$ (red line) of the single-pulse protocol (green crosses) is removed.
For $\ket{10}$ and $\ket{11}$, the infidelities are small for both protocols.
(d) Infidelity $1-\mathcal{F}$ if all atoms are initialized in $1/\sqrt{2}\left(\ket{0}+\ket{1}\right)$.
For the double-pulse protocol, the infidelity is reduced by about two orders of magnitude and increases only quadratically in $\epsilon$. 
For the single-pulse protocol, we observe a linear increase.}
\label{fig: infid}
\end{figure}
We now numerically simulate the time evolution of the system for arbitrary states of the gate atoms, taking the interaction between the gate atoms and the third one into account. 
We let the system evolve under the two pulse-protocols governed by the total Hamiltonian in Eq. \eqref{eq: general H} and calculate the fidelity 
\begin{equation}
    \label{eq: fid num}
    \mathcal{F} = \left|\braket{\Psi_{\scriptscriptstyle t}|\Psi_{\scriptscriptstyle f}}\right|^2
\end{equation}
of the effective three-qubit gate.
Here, $\ket{\Psi_{\scriptscriptstyle f}}$ is the final state after the pulse protocol has ended, assuming that all atoms were initialized in $1/\sqrt{2}\left(\ket{0} + \ket{1}\right)$. 
Without crosstalk and Rydberg excitation, the system would ideally end up in the target state $\ket{\Psi_{\scriptscriptstyle t}} = 1/2\left(\ket{00} + \ket{01} + \ket{10} - \ket{11}\right)\otimes 1/\sqrt{2}\left(\ket{0} + \ket{1}\right)$. The fidelity of the effective three-qubit gate can be seen as an extension of the two-qubit Bell state fidelity to also take into account the action on the third atom and the backaction that an excited third atom might have on the gate atoms.

The comparison of the two protocols clearly shows that using the double-pulse protocol drastically reduces the infidelity by about two orders of magnitude for experimentally relevant values of $\epsilon$.
The infidelity increases quadratically in $\epsilon$ for $\epsilon \gtrsim 10^{-3}$ and  linearly for smaller $\epsilon$-values, see Fig. \ref{fig: infid}(d). 

To understand the different regimes, we analyze the infidelity for different initial states of the gate atoms, see Fig. \ref{fig: infid}(a)-(c).
Without the double-pulse protocol, the most critical initial gate state is $\ket{00}$ as assumed, because then the Rydberg blockade does not come into play. 
Hence, the excitation of the third atom to the Rydberg state is high and we observe a linear increase in the infidelity.
With the double-pulse protocol, we can suppress this excitation and the infidelity only increases quadratically in $\epsilon$, see Fig. \ref{fig: infid}(a).
The numerically calculated values for the double-pulse protocol perfectly match the quadratic increase of $6.91\epsilon^2$ predicted by the perturbative description in Sec. \ref{sec: pert. description }.
The initial gate state $\ket{00}$ dominates the total three-qubit gate fidelity for $\epsilon \gtrsim 10^{-3}$, and we observe $1-\mathcal{F} \propto \epsilon^2$ in Fig. \ref{fig: infid}(d).

For small $\epsilon$, the initial gate state $\ket{10}$ dominates, see Fig. \ref{fig: infid}(b), leading to $1- \mathcal{F} \propto \epsilon$ for $\epsilon \lesssim \ 10^{-3}$ as seen in Fig. \ref{fig: infid}(d).
This is explained by the fact that the double-pulse protocol is not optimized for the initial gate state $\ket{10}$. 
However, because of the Rydberg blockade, the infidelity is already small for the initial state $\ket{10}$.

Because of the even stronger effect of the Rydberg blockade, the initial gate state $\ket{11}$ never dominates the total three-qubit gate fidelity.

\section{\label{sec: vdw variieren} Dependence on van der Waals interaction strength}
For simplicity, we have so far only considered an equilateral arrangement in which the distance between all three atoms and thus the van der Waals interaction strength is the same. 
To generalize this setup, the van der Waals interaction strength $V_{\scriptscriptstyle i3}$ between the third atom and the two gate atoms is now varied by increasing the distance between the third atom and the gate atoms.
\begin{figure}[t]
\centering
    \begin{subfigure}[b]{0.422\textwidth}
        \includegraphics[width=1\linewidth]{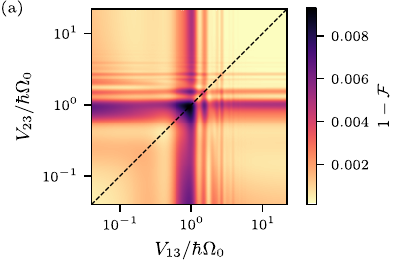}
    \end{subfigure}
    \begin{subfigure}[b]{0.48\textwidth}
        \includegraphics[width=1\linewidth]{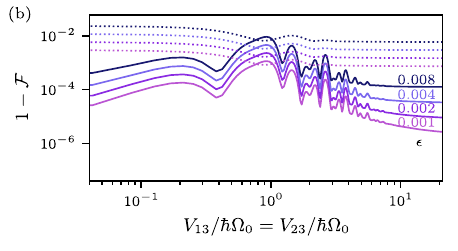}       
    \end{subfigure}
\caption{\justifying Dependence of the infidelity $1-\mathcal{F}$ on the van der Waals interaction strength between the third atom and the gate atoms.
(a) Infidelity for different interaction strengths $V_{\scriptscriptstyle i3}$ for $\epsilon = 0.008 $.
(b) Infidelity along the dashed line in (a) for $V_{\scriptscriptstyle 13} = V_{\scriptscriptstyle 23}$ and different $\epsilon$-values, when applying the double-pulse protocol. For comparison, the infidelity is plotted for the single-pulse protocol (dotted lines) for the same $\epsilon$-values. The infidelity exhibits the same behavior for all $\epsilon$-values. }
\label{fig: vdW}
\end{figure}
There are two regimes in which the infidelity is suppressed, see Fig. \ref{fig: vdW}.

For large interactions $V_{\scriptscriptstyle i3}$, we are in the regime that we discussed in the prior section, and the double-pulse protocol suppresses the infidelity by about two orders of magnitude.

Weak interactions lead to a negligible Rydberg blockade. Hence, the double-pulse protocol works independently of the initial state of the gate atoms and the infidelity is suppressed by an amount similar to that in the previous regime.

In contrast, for $V_{\scriptscriptstyle i3}/\hbar\Omega_{\scriptscriptstyle 0} \approx 1$, the infidelity cannot be suppressed by the double-pulse protocol. While the protocol still works for the gate atoms being initialized in $\ket{00}$, we have strong contributions of $\ket{10}$ and $\ket{11}$ to the infidelity. For these initial states, the Rydberg blockade is neither weak enough to be able to use the double-pulse protocol nor strong enough to suppress the unwanted excitation of Rydberg states. Hence, it is advisable to avoid this region for experimental realizations.

For lattice geometries avoiding this intermediate-interaction regime, the double-pulse protocol suppresses the leading contribution to the amplitude error also if multiple atoms are affected by crosstalk.

\section{\label{sec: phasen variieren} Cancellation of remaining phase errors}
\begin{figure}[t]
\centering
    \begin{subfigure}[b]{0.477\textwidth}
        \includegraphics[width=1\linewidth]{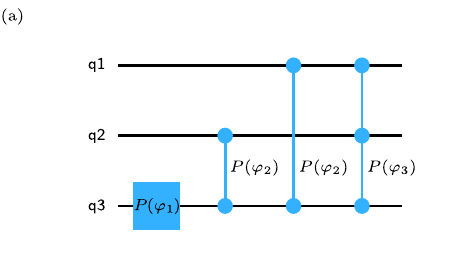}
    \end{subfigure}
    \begin{subfigure}[b]{0.48\textwidth}
        \includegraphics[width=1\linewidth]{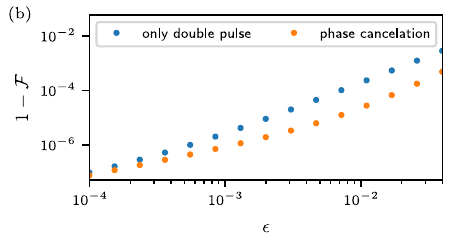}
    \end{subfigure}
\caption{\justifying Reducing the infidelity by canceling remaining phase errors.
(a) Quantum circuit for canceling the phase errors.
(b) By applying the circuit after the double-pulse protocol, the infidelity $1-\mathcal{F}$ is reduced by almost an additional order of magnitude.}
\label{fig: cancel phase}
\end{figure}
The double-pulse protocol reduces the amplitude error and thus the leakage out of the qubit space. The remaining errors are mainly phase errors. These errors are coherent errors within the qubit space. Therefore, they can be compensated with gates as described in the following.

If the third atom is initialized in $\ket{0}$, it will stay in this state.
The gate atoms then evolve according to Eq. \eqref{eq: mapping CZ gate}.
When the third atom is initialized in state $\ket{1}$, it can collect a phase depending on the initial state of the gate atoms. 
Then, the following operation is performed:
\begin{align}
\label{eq: mapping phase}
    \begin{split}
        &\ket{001} \rightarrow \ket{001}e^{i\phi_{\scriptscriptstyle 1}} = \ket{001}e^{i\varphi_{\scriptscriptstyle 1}}\\
        &\ket{011} \rightarrow \ket{011} e^{i\phi_{\scriptscriptstyle 2}} = \ket{011} e^{i(\varphi_{\scriptscriptstyle 1} + \varphi_{\scriptscriptstyle 2})}\\
        &\ket{101} \rightarrow \ket{101} e^{i\phi_{\scriptscriptstyle 2}}
        = \ket{101} e^{i(\varphi_{\scriptscriptstyle 1} + \varphi_{\scriptscriptstyle 2})}\\
        &\ket{111} \rightarrow \ket{111} e^{i(\phi_{\scriptscriptstyle 3}-\pi)} = \ket{111} e^{i(3\varphi_{\scriptscriptstyle 1} + 2\varphi_{\scriptscriptstyle 2} + \varphi_{\scriptscriptstyle 3}-\pi)}\\
        & \ket{**0} \rightarrow \text{see Eq. \eqref{eq: mapping CZ gate}}.
    \end{split}
\end{align}
The phases $\phi_{\scriptscriptstyle 1}$, $\phi_{\scriptscriptstyle 2}$ and $\phi_{\scriptscriptstyle 3}$ can be canceled using the quantum circuit in Fig. \ref{fig: cancel phase}(a).
The phases $\varphi_{\scriptscriptstyle 1}, \varphi_{\scriptscriptstyle 2}$ and $\varphi_{\scriptscriptstyle 3}$ are the phases of the single-qubit phase gate, the two two-qubit controlled-phase gates and the three-qubit controlled-phase gate as shown in Fig. \ref{fig: cancel phase}(a). 
In the case that the third atom was not equidistant to the two gate atoms, two different controlled-phase gates would be required in the circuit. 

If the circuit is applied in addition to the double-pulse protocol, the fidelity can further be reduced by nearly an order of magnitude for $\epsilon > 10^{-3}$ as shown in Fig. \ref{fig: cancel phase}(b). 
Note that, in principle, the additional two-qubit gates themselves introduce new crosstalk. However, due to their small phases, the gates only require a small Rydberg population and the error from the crosstalk is smaller than the remaining infidelity.

\section{Conclusion and Outlook}
We have presented a method to suppress crosstalk in neutral atom quantum processors that use local addressing to implement controlled-Z gates. In these systems, a fraction of the laser light that is applied locally to implement gates typically leaks to other atoms. We have studied the consequences of the crosstalk, using a setup comprising two gate atoms and a neighboring third atom. We have shown that the main effect of the leaking laser light is an amplitude error on the third atom. Using perturbation theory, we have constructed a gate protocol that uses two pulses to cancel this error in leading order. Numerical simulations confirm that the double-pulse protocol reduces the infidelity by two orders of magnitude across a broad, experimentally relevant range of laser intensities and van der Waals interaction strengths. To suppress remaining phase errors, we have proposed a quantum circuit to cancel them, reducing the infidelity by almost another order of magnitude.

Future research could focus on developing compilers that automatically add the additional gates to cancel phase errors. While these additional gates might be implementable with high fidelity because these gates only apply tiny phases to the qubit state, a reduction of the gate count could be achieved by an optimizing compiler.

\section{Acknowledgements}
We thank the MUNIQC-Atoms team and the \textsc{Quantum Länd} team for useful experimental insights. We acknowledge funding from the Federal Ministry of Research, Technology and Space (BMFTR) under the grants QRydDemo and MUNIQC-Atoms, and from the Horizon Europe programme HORIZON-CL4-2021-DIGITALEMERGING-01-30 via the project 101070144 (EuRyQa). 
S.W. and H.P.B. acknowledge funding from the Deutsche Forschungsgemeinschaft (DFG) under the Priority Programme SPP 2514, and F.M. from the Federal Ministry of Education and Research under the grant CiRQus.

\bibliographystyle{apsrev4-2-author-truncate}
\bibliography{manuscript}

\end{document}